\documentstyle[prl,aps,twocolumn,epsf]{revtex}
\begin{document}
\draft

%\begin{multicols}{2}

{\bf Comment on ``Search for two-scale localization in disordered wires in a
magnetic field''}

%\maketitle

In a recent work Schomerus and Beenakker\cite{Been-Sch} tried to check
numerically our prediction \cite{WE} about a two-scale localization in
disordered wires in a weak magnetic field. According to our theory an
exponential decay of wave functions with the localization length
$L_{\rm co}$ of the orthogonal ensemble is followed at distances
$x>x_{H}\sim L_c\ln X^{-1}$ by another exponential decay with the
length $L_{\rm cu}=2L_{\rm co}$ of the unitary ensemble ($L_{c}$ is 
either $L_{\rm co}$ or $L_{\rm cu}$). Here $X$ is
proportional to the magnetic field. In our case $X\ll 1$; the crossover
between the ensembles occurs when $X\sim 1$.

Studying the transmittance the authors of \cite{Been-Sch} did not confirm
our theory. The most probable explanation they suggested is that
only rare states could have the two scales while ``typical'' states 
have the only length $L_{\rm co}$.

We would like to suggest another explanation of the discrepancy,
namely, that the Borland conjecture relating the transmittance and wave
functions to each other may not be used for studying the effect of the
magnetic field on the tails of wave functions. The physical reason is
that the wave functions near the ends of the sample determining the
transmittance are distorted by the presence of the metallic leads.

Although the level smearing is always present near the ends, its influence
is not very important for calculation of the main body of wave functions
decaying with the localization length $L_{\rm co}$. This is due to the
fact that the smearing decays also exponentially and becomes smaller than
the relevant energy $E_{c}=D_{0}/L_{c}^{2}$ already at the distance $x\sim
L_{c}$ from the lead, where $D_{0}$ is the diffusion coefficient. So, one
can expect a change of the coefficient $A$ in a wave function of the type
$\psi \sim A\exp \left( -L/L_{c}\right) $ but not of the exponent ($L$ is
the length of the sample). This may justify the use of the Borland
conjecture for the ``pure'' (orthogonal or unitary) ensembles.

However, the smearing does influence the tails of the wave functions. The
tails decaying with the localization length $L_{\rm cu}$ can be observed
only if the smearing is smaller than the energy $E_{H}=X^{2}E_{c}\ll
E_{c}$, which corresponds to  $x\gg x_{H}$. At smaller distances the
information about the original states is lost.

A detailed consideration is given elsewhere \cite{WE1} but here we present
only the main ideas. Within the supersymmetry technique the transmittance
can be written in a form \cite{Zirn,book} $T=\int {\rm d}\,Q_{1}$ ${\rm d}%
\,Q_{2}\,$ $\Psi _{1}\,Q_{1}\,\Gamma (x_{1},x_{2};Q_{1},Q_{2})\,Q_{2}\Psi
_{2}$, where certain elements of the $Q$-matrix are taken at the ends of the 
$x_{1}$, $x_{2}$ of the wire and  $\Gamma $ plays the role of a Green
function. In contrast to a standard expression for infinite wires \cite{book}%
, the function $\Psi $ does not obey an $X$-dependent differential equation
but is given by $\Psi (Q)=\gamma \exp \left[ -\gamma ({\rm Str}\Lambda Q)%
\right] $, where $\gamma $ is a large number.

Instead of correlators themselves, we calculated \cite{WE} their derivatives
in $X$, %This is important because, For the derivatives, 
ensuring the main contribution to come from large values $\lambda _{1c}$, $%
\lambda _{1d}$ of radial variables of the $Q$-matrix. However, repeating the
same for the conductance of the open wire we see that the derivative of $%
\Psi $ vanishes. The $Q$-matrix near the lead is $Q\simeq \Lambda $, such
that $\lambda _{1c}$ $\simeq 1$ and, in contrast to \cite{WE}, we do not get
any anomalous contribution from $\lambda _{1c}\sim 1/X.$ So, a weak magnetic
field $X\ll 1$ does not influence the transmittance and noticeable changes
can occur at $X\sim 1$ only.

We come to the same conclusion using arguments similar to those of Ref. \cite
{Mott}. In wires all states are localized and we can speak of the wire in
terms of a chain of grains of the size $L_{c}$ with the energy separation in
an isolated grain $E_{c}\sim D_{0}/L_{c}^{2}$. Then, we see that the
crossover between the ensembles occurs when a characteristic energy,
$E_{H}=X^{2}D_{0}/L_{c}^{2}$ becomes comparable with $E_{c}$.

Nevertheless, taking into account an overlap between the grains one can find
at $X\ll 1$ states separated by small energies within the window $E_{H}$.
They cannot be localized at distances of order of $L_{c}$ because they would
hybridize and the corresponding splitting would be of the order of $E_{c}$.
However, if two states are separated by a large distance $x\gg L_{c}$, the
splitting energy $\Delta _{x}$ decays exponentially $\Delta _{x}\sim
E_{c}\exp (-x/L_{c})$. Comparing this energy with $E_{H}$, we immediately
obtain the distances $x_{H}\sim L_{c}\ln 1/X$ at which the states of
isolated grains are mixed and matrix elements of the vector potential
between the states are not exponentially small. At distances exceeding $x_{H}
$ the states in the energy window $E_{H}$ are well mixed by the magnetic
field and their correlations should decay with the localization length $%
L_{\rm cu}$. At the same time, any level smearing exceeding $E_{H}$ allows
us to neglect the magnetic field, which means that the wave functions near
the leads are not influenced by it. Therefore, the transmittance is also
insensitive to a weak magnetic field and the numerical results of Ref. \cite
{Been-Sch} cannot be used as a check of our predictions \cite{WE}.

\bigskip \noindent
A.V.~Kolesnikov$^{1}$ and K.B.~Efetov$^{1,2}$\newline
$^{1}$Fakult\"{a}t f\"{u}r Physik und Astronomie, Ruhr-Universit\"{a}t%
\newline
Bochum, Universit\"{a}tsstr. 150, Bochum, Germany\newline
$^{2}$L.D. Landau Inst. for Theor. Phys., Moscow, Russia


%\end{multicols}


\begin{references}
\bibitem{Been-Sch}  H.~Schomerus and C.W.J.~Beenakker, \prl {\bf 84}, 3927
(2000).

\bibitem{WE}  A.V.~Kolesnikov, K.B.~Efetov, \prl {\bf 83}, 3689 (1999).

\bibitem{WE1}  A.V.~Kolesnikov and K.B.~Efetov, to be published.

\bibitem{Zirn}  M.R.~Zirnbauer, \prl {\bf 69}, 1584 (1992).

\bibitem{book}  K.B.~Efetov, {\it Supersymmetry in Disorder and Chaos }
(Cambridge University Press, New York 1997).

\bibitem{Mott}  N.F.~Mott and M.~Kaveh, Adv. Phys. {\bf 34}, 329 (1985).
\end{references}
\end{document}